\documentstyle[osa,manuscript,epsf,tighten]{revtex}  % DON'T CHANGE
\begin{document}                % INITIALIZE - DONT CHANGE
\title{Laser Excitation of Polarization Waves in a Frozen Gas}
\author{V. Celli and J. S. Frasier}
\address{Department of Physics, University of Virginia \\
P. O. Box 400714 \\
Charlottesville, VA 22904-4714}
\maketitle
\begin{abstract}
Laser experiments with optically excited frozen gases entail the
excitation of polarization waves. In a continuum approximation the
waves are dispersionless, but their frequency depends on the
direction of the propagation vector. An outline is given of the
theory of transient phenomena that involve the excitation of these
waves by a resonant dipole-dipole transfer process.
\end{abstract}

\section{INTRODUCTION}

\label{intro}

If the energy associated with the translational temperature of a
gas is low compared to the relevant excitation energies and to the
energy associated with the inverse observation time, then the gas
can be considered as ``frozen'' and is analogous to an amorphous
solid, or a glass. Such a frozen gas can be produced by laser
cooling, and the atoms or molecules can be selectively placed in
certain excited states. When only a few states are involved, the
system is analogous to a spin glass and can support polarization
waves (dipolar Frenkel excitons) that are analogous to spin waves.
In this paper, we introduce a basic example of such waves and
discuss their excitation.

In laser experiments, the frozen gas is set up in a
non-equilibrium electronic state, and the evolution of this state
is measured after some elapsed time. Relaxation processes can be
neglected for the duration of the experiment. Thus, the electronic
state evolves towards dynamical equilibrium through the excitation
of polarization waves in the random medium of the frozen gas. The
theory of this process must determine the spectrum of the
polarization waves as well as their excitation probability during
the transient evolution to dynamical equilibrium. The two aspects
of the problem will often be intermingled. However, we have chosen
to work on a system where they can be separated to some extent.

The system under consideration is fully described in an earlier
paper by J. S. Frasier, V. Celli, and T. Blum,\cite{prapaper}
henceforth to be referred to as FCB. In the present paper, we
develop an improved theory of the particular case that is
discussed in Sec.~III of FCB, under the heading ``Sparse
$ss^{\prime}\rightarrow pp^{\prime}$ in a bath of $sp\rightarrow
ps$.'' Simply put, we consider one atom initially in state
$s^{\prime}$ that is immersed in a frozen gas of atoms in state
$s$. The $s^{\prime}$ state is connected to a $p^{\prime}$ state
by a dipole matrix element $\mu^{\prime}$, and an $s$ state is
connected to a $p$ state by a dipole matrix element $\mu$. The $p$
and $p^{\prime}$ states are singled out from all possible states,
because the $ss^{\prime}\rightarrow pp^{\prime}$ transition is
resonant, i.e., $\epsilon^{\prime}_{s}+\epsilon_{s} =
\epsilon^{\prime}_{p}+\epsilon_{p}$, and it is known that the
transition rate peaks sharply at
resonance.\cite{mourachko,anderson} Deviations from resonance are
not considered in this paper. The resonance condition is achieved
by the application of a static electric field. The
$ss^{\prime}\rightarrow pp^{\prime}$ process excites polarization
waves described by $sp\rightarrow ps$ transitions in the gas of
$s$ atoms. Typically, the number $N$ of $s$ atoms in the volume
$\Omega$ is such that $N/\Omega\simeq 10^{9}$
$\mathrm{atoms/cm^{3}}$. The $s^{\prime}$ atoms are ``sparse'' if
their number $N^{\prime}$ is such that $N^{\prime}/N\ll 1$, and in
that case the result for $N^{\prime}$ atoms can be obtained by
computing the result for one $s^{\prime}$ atom, averaging over
distributions of the $s$ atoms, and then multiplying this result
by $N^{\prime}$.

\section{THE EQUATIONS OF TIME EVOLUTION}

\label{eqtimeevol}

We begin with the set of coupled differential equations in
Eq.~(24) of FCB, Fourier transformed in time:
% Create a counter to go through a, b, c.
\newcounter{tlet}
% Set the counter to one.
\setcounter{tlet}{1}
% Then change the command \theequation to display
% the counter differently.
\renewcommand{\theequation}{\arabic{equation}\alph{tlet}}
\begin{eqnarray}
\label{dilute-eqns}
 \omega a_{0} &=& \sum_{l}V_{l}c_{l}+i, \\
\stepcounter{tlet}\addtocounter{equation}{-1}\label{a0dot}
 \omega c_{l} &=& V_{l}a_{0} + \sum_{m}U_{lm}c_{m}.
\label{cldot}
\end{eqnarray}
\renewcommand{\theequation}{\arabic{equation}}
As in FCB, $a_{0}\left( t \right)$ is the amplitude of the state
in which the atom at the origin is in state $s^{\prime}$ and all
other atoms are in state $s$, while $c_{l}\left( t \right)$ is the
amplitude of the state in which the atom at the origin is in state
$p^{\prime}$, the atom at position ${\mathbf{r}} _{l}$ is in state
$p$, and all other atoms are in state $s$. The quantity $V_{l}$ is
the interaction potential
\begin{equation}
V\left( \mathbf{r} \right) = \frac{\mu\mu^{\prime}}{r^{3}}\left(
1-3\cos^{2}\theta_{r}\right) =
-2\frac{\mu\mu^{\prime}}{r^{3}}P_{2}\left( \cos\theta_{r}\right),
\label{V}
\end{equation}
evaluated at ${\mathbf{r}} = {\mathbf{r}} _{l}$, and $\theta_{r}$
is the polar angle. Similarly, $U_{lm}$ is
\begin{equation}
U\left( \mathbf{r}\right) = \frac{\mu^{2}}{r^{3}}\left(
1-3\cos^{2}\theta_{r} \right) = -2\frac{\mu^{2}}{r^{3}}P_{2}\left(
\cos\theta_{r}\right), \label{U}
\end{equation}
evaluated at ${\mathbf{r}} = {\mathbf{r}} _{l} - {\mathbf{r}}
_{m}$. The angular factor present in $V$ and $U$ has been chosen
for mathematical convenience: the actual angular factor is more
complicated, as discussed in Sec.~\ref{conc}. In FCB, all the
results depend only on an angular average of $V^{2}$ and, as is
shown in the Appendix of that paper, one can work with the
effective angle-averaged potential $V\left( r\right) = \left(
4/3\sqrt{3}\right) \mu\mu^{\prime}/r^{3}$. In the treatment
presented here, however, the angular dependence plays a more
important role.

The potentials of Eqs.~(\ref{V}) and (\ref{U}) have particularly
simple Fourier transforms. Defining
\begin{equation}
V\left(\mathbf{k}\right) = \int
d^{3}r\,V\left(\mathbf{r}\right)e^{-i\mathbf{k}\cdot\mathbf{r}},
\end{equation}
we find, using the standard expansion of
$e^{-i\mathbf{k}\cdot\mathbf{r}}$ in spherical Bessel functions
and spherical harmonics,\cite{footnote2}
\begin{eqnarray}
V\left(\mathbf{k}\right) &=& -8\pi\mu\mu^{\prime}
P_{2}\left(\cos\theta_{k}\right)\int_{0}^{\infty}\frac{dr}{r}j_{2}
\left( kr\right) \nonumber \\ \label{Vk}&=&
-\frac{8\pi}{3}\mu\mu^{\prime} P_{2}\left(\cos\theta_{k}\right).
\end{eqnarray}
Similarly,
\begin{equation}
U\left(\mathbf{k}\right) = -\frac{8\pi}{3}\mu^{2}
P_{2}\left(\cos\theta_{k}\right). \label{Uk}
\end{equation}

With these expressions we can obtain explicit solutions for the
polarization waves in the continuum approximation, and we can then
use these approximate solutions to describe the dynamics of a
frozen gas state.

\section{THE CONTINUUM MODEL}

\label{contmodel}

We change the sums in Eqs.~(1) to integrations over space, so that
the evolution equations become
% Set the counter to one.
\setcounter{tlet}{1}
% Then change the command \theequation to display
% the counter differently.
\renewcommand{\theequation}{\arabic{equation}\alph{tlet}}
\begin{eqnarray}
\label{dilute-eqns-int} \omega a_{0} &=& \frac{N}{\Omega}\int
d^{3}r\,V\left(\mathbf{r}\right)c\left(\mathbf{r}\right)+i, \\
\stepcounter{tlet}\addtocounter{equation}{-1}\label{a0dotint}
\omega c\left(\mathbf{r}\right) &=& V\left( {\mathbf{r}}
\right)a_{0} + \frac{N}{\Omega}\int
d^{3}r^{\prime}\,U\left(\mathbf{r} -
\mathbf{r}^{\prime}\right)c\left(\mathbf{r}^{\prime}\right),
\label{crdotint}
\end{eqnarray}
\renewcommand{\theequation}{\arabic{equation}}
where $c\left(\mathbf{r}\right)$ is now an averaged quantity.
Fourier transforming, we obtain simply
% Set the counter to one.
\setcounter{tlet}{1}
% Then change the command \theequation to display
% the counter differently.
\renewcommand{\theequation}{\arabic{equation}\alph{tlet}}
\begin{eqnarray}
\label{dilute-eqns-int-ft} \omega a_{0} &=& \frac{N}{\Omega}\int
\frac{d^{3}k}{\left( 2\pi\right)^{3}}\,V\left(
\mathbf{k}\right)c\left( \mathbf{k}\right) + i
\\
\stepcounter{tlet}\addtocounter{equation}{-1} \omega
c\left(\mathbf{k}\right) &=&
V\left(\mathbf{k}\right)a_{0}+\frac{N}{\Omega}
U\left(\mathbf{k}\right)c\left(\mathbf{k}\right).
\end{eqnarray}
\renewcommand{\theequation}{\arabic{equation}}

\begin{figure}[tbp]
\begin{center}
\leavevmode \hbox{\epsffile{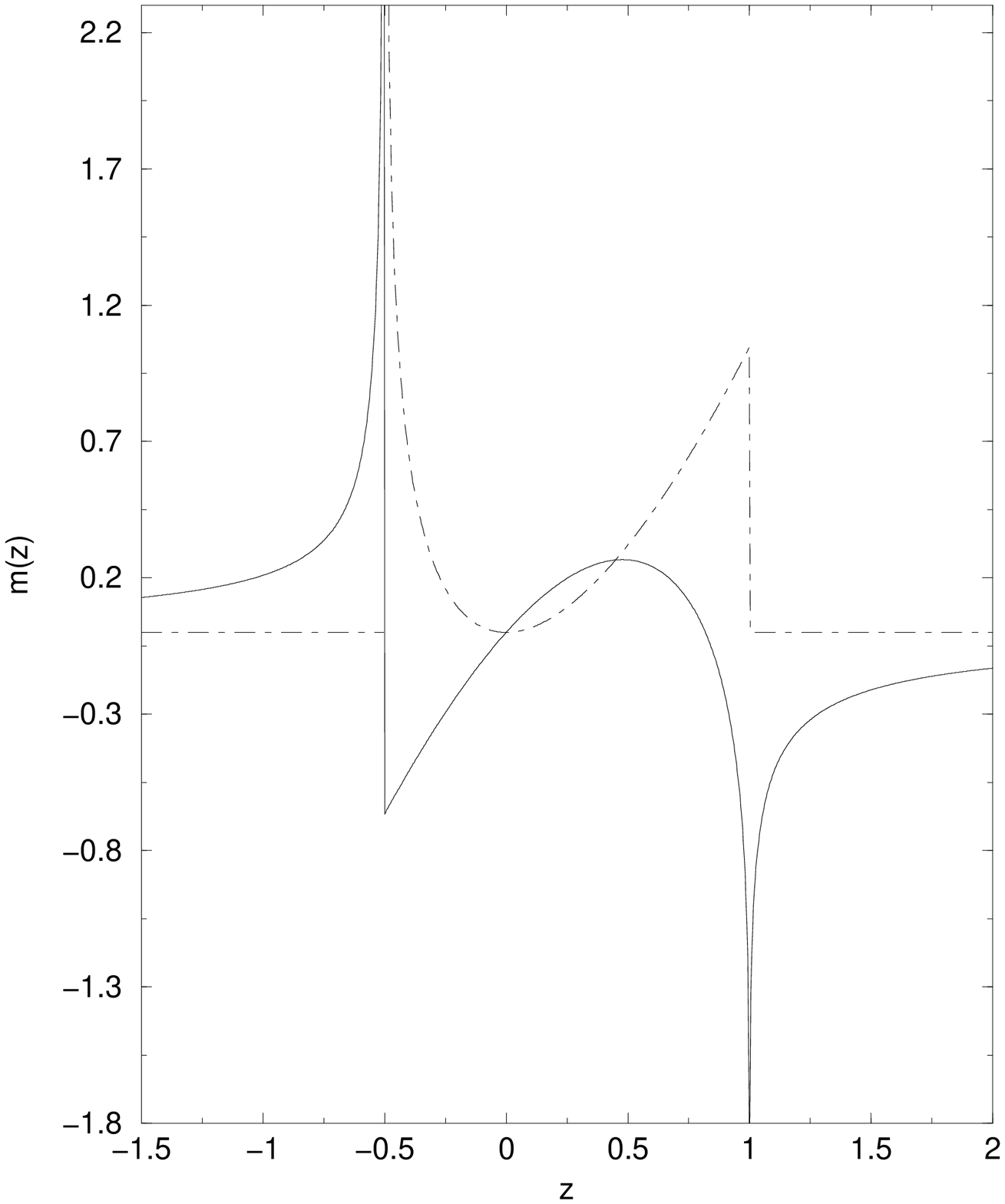}}
\end{center}
\caption{ The real part (solid line) and imaginary part (dashed
line) of the function $m\left( z\right)$ for $z$ just above the
real axis. } \label{figure1}
\end{figure}

The last equation shows that the frequencies of the polarization
waves are
\begin{equation}
\omega\left( \mathbf{k}\right) = \frac{N}{\Omega}U\left(
\mathbf{k}\right), \label{omegak}
\end{equation}
independent of the magnitude of $\mathbf{k}$, but dependent on the
direction, according to Eq.~(\ref{Uk}). This peculiar dispersion
relation, which is analogous to well-known results in spin wave
theory, is a consequence of the long range nature of the
dipole-dipole interaction. We expect it to be useful for values of
$k$ up to the order of the inverse of the interatomic
spacing\@.\cite{cohen} The solution of Eqs.~(8) gives
\begin{equation}
a_{0}\left(\omega\right) = \frac{i}{\omega -
M\left(\omega\right)}, \label{a0M}
\end{equation}
where
\begin{equation}
M\left(\omega\right) = \frac{N}{\Omega}\int\frac{d^{3}k}{\left(
2\pi\right)^{3}} \frac{V\left(\mathbf{k}\right)^{2}}{\omega -
\omega\left(\mathbf{k}\right)}. \label{M}
\end{equation}
Carrying out the angular integration, we find
\begin{equation}
M\left(\omega\right) =
-\frac{4}{3\pi}\left(\mu^{\prime}\right)^{2}
\left[\int_{0}^{k_{c}}dk\,k^{2}\right]m\left( z\right),
\label{Meqn}
\end{equation}
where, leaving for Sec.~\ref{capproxrandmed} a discussion of the
effective cutoff $k_{c}$,
\begin{equation}
m\left( z\right) = z+\frac{z^{2}}{\sqrt{3+6z}} \log\left(
\frac{1-\sqrt{\frac{3}{1+2z}}}{1+\sqrt{\frac{3}{1+2z}}} \right),
\end{equation}
and $z = \omega/u_{0}$ with
\begin{equation}
u_{0} = \frac{8\pi}{3}\mu^{2}\frac{N}{\Omega}.
\end{equation}
A plot of ${\mathrm{Re}}\, m$ and ${\mathrm{Im}}\, m$ on the real
axis is shown in Fig.~\ref{figure1}. For large $z$,
\begin{equation}
m\left( z\right) \sim -\frac{1}{5z} + O\left(
\frac{1}{z^{2}}\right). \label{Mexp}
\end{equation}

As expected, $m\left( z\right)$ is an analytic function in the
upper $z$-plane, with a cut extending from $z = -1/2$ to $z = 1$.
On the real axis outside the cut $m\left( z\right)$ is real, and
the spectral density on the cut is given by
\begin{equation}
{\mathrm{Im}} \left[ m\left( z \right) \right] =
\frac{\pi}{\sqrt{6}}\frac{z^{2}}{\sqrt{z+\frac{1}{2}}}.
\end{equation}

\section{CONTINUUM APPROXIMATION FOR THE RANDOM MEDIUM}

\label{capproxrandmed}

To complete the evaluation of the self-energy
$M\left(\omega\right)$ of Eq.~(\ref{Meqn}), it is necessary to
introduce a cutoff in the $k$ integral. We can do this in a way
that makes the result applicable, at least approximately, to a
random medium.

We know from Sec.~II of FCB that the solution of Eqs.~(1) for $U =
0$ is of the form of Eq.~(\ref{a0M}), with
\begin{equation}
M\left( \omega\right)_{U=0} = \frac{1}{\omega}\sum_{l}V_{l}^{2},
\label{MsecIIfcb}
\end{equation}
while in the continuum limit, using
Eq.~(\ref{Vk}),\cite{footnote1}
\begin{eqnarray}
M\left( \omega\right)_{U=0} &=&
\frac{N}{\Omega}\int\frac{d^{3}k}{\left(
2\pi\right)^{3}}\,\frac{V\left({\mathbf{k}} \right)^{2}}{\omega}
\nonumber \\ &=&
\frac{32}{45}\frac{N}{\Omega}\left[\int_{0}^{k_{c}}
dk\,k^{2}\right]\frac{\left(\mu\mu^{\prime}\right)^{2}}{\omega}.
\end{eqnarray}
Therefore, we see that if we make the substitution
\begin{equation}
\frac{N}{\Omega}\int_{0}^{k_{c}}dk\,k^{2} \rightarrow
\frac{45}{8}\sum_{l}\frac{1}{r_{l}^{6}}\left[
P_{2}\left(\cos\theta_{r_{l}}\right)\right]^{2},
\end{equation}
then in the limit of $U$ going to zero our continuum result will
agree with the result of FCB. Applying this substitution, we find
that Eq.~(\ref{Meqn}) becomes
\begin{eqnarray}
M\left( \omega\right) &=&
-\frac{15}{2\pi}\frac{\Omega}{N}\left(\mu^{\prime}\right)^{2}
m\left( z\right) \sum_{l}\frac{1}{r_{l}^{6}}\left[
P_{2}\left(\cos\theta_{r_{l}}\right)\right]^{2} \nonumber \\ &=&
-\frac{5m\left( z\right)}{u_{0}}\sum_{l}V_{l}^{2}. \label{newMeqn}
\end{eqnarray}

We can now average over atom positions by the technique described
in FCB, Eq.~(9), with the result that, for ${\mathrm{Im}}\, z >
0$,
\begin{eqnarray}
\left( a_{0}\right)_{av} &=& \int_{0}^{\infty}d\beta\, \exp \left(
i\beta\omega - v\sqrt{-\frac{5i\beta}{u_{0}} m\left( z\right)}
\right) \nonumber \\ &=& \int_{0}^{\infty}\frac{d\beta}{u_{0}}
\exp\left( i\beta z -  \frac{v}{u_{0}}\sqrt{-5i\beta m(z)}
\right), \label{a0av}
\end{eqnarray}
where, as in FCB Eq.~(A2),
\begin{equation}
v=\frac{16\pi^{3/2}}{9\sqrt{3}}\,\frac{N}{\Omega}\mu\mu^{\prime },
\label{v}
\end{equation}
so that
\begin{equation}
\frac{v}{u_{0}}=\frac{2\sqrt{3\pi}}{9}\frac{\mu^{\prime}}{\mu}.
\end{equation}
The integral in Eq.~(\ref{a0av}) can be expressed in terms of the
complementary error function.\cite{prapaper}

To develop approximations that automatically include a momentum
cutoff we introduce the Green's function of Eq.~(1b), which
satisfies the equation
\begin{equation}
\omega G_{ln} = \delta_{ln} + \sum_{m}U_{lm}G_{mn}.
\end{equation}
Then the solution of Eqs.~(1) is of the form of Eq.~(\ref{a0M}),
with the exact self-energy given by
\begin{equation}
M\left( \omega\right) = \sum_{lm}V_{l}G_{lm}V_{m}.
\end{equation}

Similarly, the on-site Green's function, which clearly cannot be
treated in the continuum approximation, can be written as
\begin{equation}
G_{nn} = \frac{1}{\omega -\sum_{mp}U_{nm}G_{mp,[n]}U_{pn}},
\label{Gnn}
\end{equation}
while, for $l\neq n$,
\begin{equation}
G_{ln}=G_{ll,[n]}U_{ln}G_{nn}+\sum_{m\neq
l}G_{lm,[n]}U_{mn}G_{nn},
\end{equation}
where $G_{lm,[n]}$ is the Green's function for a medium where the
$n$ atom is absent. These equations are exact, but we now replace
all occurrences of $G_{nn}$ on the right hand side by
$\left\langle G\right\rangle $, where $\left\langle G\right\rangle
$ denotes the average value of $G_{nn}$, which does not depend on
$n$. Averaging and Fourier transforming the last equation and
setting
\begin{equation}
\left\langle G_{lm}\right\rangle =
\int\frac{d^{3}k}{\left(2\pi\right)^{3}}G\left( {\bf k}\right)\exp
\left[ i{\bf k\cdot }\left( {\bf r}_{l}-{\bf r}_{m}\right) \right]
\end{equation}
we obtain
\begin{eqnarray}
G({\bf k}) &=& \frac{\left\langle G\right\rangle ^{2}U({\bf
k})}{1-\left(N/\Omega \right)\left\langle G\right\rangle U({\bf
k})} \\ \nonumber &=& \frac{\Omega}{N}\left(\frac{1}{\left\langle
G\right\rangle ^{-1}-(N/\Omega )U({\bf k})}-\left\langle
G\right\rangle \right).
\end{eqnarray}
Similarly, writing the approximate version of Eq.~(\ref{Gnn}) in
the form
\begin{eqnarray}
G_{nn} &=& \frac{1}{i}\int_{0}^{\infty}d\beta\,\exp\left(
i\beta\omega - i\beta\sum_{m}U_{nm}^{2}\left\langle G\right\rangle
\right) \\ \nonumber &\times &\exp\left( -i\beta\sum_{m\neq
p}U_{nm}G_{mp,\left[ n \right]}U_{pn} \right),
\end{eqnarray}
and expanding the second exponential, we obtain the
self-consistent equation
\begin{eqnarray}
\left\langle G\right\rangle &=& \frac{1}{i}\int_{0}^{\infty
}d\beta\,\exp \left( i\beta \omega +\left( -i\beta \left\langle
G\right\rangle \right) ^{1/2}u\right) \\ \nonumber &\times &\left(
1 - 2i\beta\int\frac{d^{3}k}{\left(2\pi \right)^{3}}U_{eff}({\bf
k})^{2}G\left({\bf k}\right) + \dots \right), \label{selfcon}
\end{eqnarray}
where $U_{eff}\left({\bf k}\right)$ is the Fourier transform of
\begin{equation}
U_{eff}({\bf r}) = U({\bf r})\exp\left( -i\beta \left\langle
G\right\rangle U^{2}({\bf r})\right).
\end{equation}
This shows how a cutoff is automatically introduced into the
theory. With these values of $\left\langle G\right\rangle $ and
$\left\langle G_{lm}\right\rangle $ we have approximately
\begin{equation}
M\left( \omega \right) =\sum_{l}V_{l}^{2}\left\langle
G\right\rangle +\sum_{l\neq n}V_{l}\left\langle
G_{ln}\right\rangle V_{n}. \label{M0best}
\end{equation}

The self-consistency equation (\ref{selfcon}) has been discussed
extensively in the approximation where the contribution of
$G\left( {\mathbf{k}}\right)$ is neglected (i.e., where contents
of the last set of parentheses in Eq.~(\ref{selfcon}) are replaced
by unity).\cite{wolynes} It is believed that this approximation is
adequate when the polarization waves are localized in the random
medium.

If $\left\langle G\right\rangle $ is known as a function of
$\omega$, one can obtain $\left\langle a_{0}\right\rangle $
directly. Comparing Eq.~(\ref {Gnn}) with
\begin{equation}
a_{0}\left( \omega \right) =\frac{i}{\omega
-\sum_{lm}V_{l}G_{lm}V_{m}},
\end{equation}
we see that for an infinite medium
\begin{equation}
\left\langle a_{0}\left( \omega \right) \right\rangle =i\left(
\frac{\mu}{\mu^{\prime}}\right)^{2}\left\langle G\left( \omega
\left( \frac{\mu}{\mu^{\prime}}\right)^{2}\right) \right\rangle.
\end{equation}

It is also possible, in principle, to compute the average of
$\left|a_{0}\left( t\right)\right|^{2}$ for the frozen gas. The
result for $a_{0}\left(\omega\right)$ with the
$M\left(\omega\right)$ of Eq.~(\ref{newMeqn}) can be inverse
Fourier transformed and squared, and then the result must be
averaged over all ${\mathbf{r}}$. We have already discussed the
cut of $m\left( z\right)$ in Sec.~\ref{contmodel}, and one can see
from Eq.~(\ref{a0M}) and the plots of Fig.~\ref{figure1} that
$a_{0}\left( t\right)$ will have a pole at some $z < -1/2$, a pole
at $z = 0$, and a pole at some $z > 1$. Therefore the inverse
Fourier transform can be computed by summing the residues of these
poles and integrating around the cut, a process which is amenable
to numerical techniques. However, the result is unreliable because
the pole at $z = 0$ is not really present in a random medium. (In
the continuum approximation, waves of zero frequency travel in the
direction where $P_{2}\left( \cos\theta\right) = 0$ and are not
excited by the coupling $V$, which vanishes at this angle.) A
similar calculation using the result of Eq.~(\ref{M0best}) is very
cumbersome. In practice, we have found it preferable to solve the
equations (1) numerically for a given configuration of a few
hundred atoms, and then average over an ensemble of
configurations. These results will be reported in a separate
publication.

\section{DISCUSSION AND CONCLUSIONS}

\label{conc}

In the course of our theoretical investigation of the transient
behavior of an optically excited gas, we have come across a
particular class of polarization waves. They are Frenkel dipolar
excitons, with the difference that the degeneracies of the $s$ and
$p$ states have been lifted by the application of a static
magnetic field. First, we note that the spin-orbit coupling
separates the $p_{1/2}$ states from the $p_{3/2}$ states. Then,
considering the $p_{1/2}$ states, we see that the interaction
between the $\left| s,\,\uparrow\right\rangle$ and $\left|
p_{1/2},\, m_{J} = \frac{1}{2}\right\rangle$ states is indeed
proportional to $P_{2}\left( \cos\theta\right)$, where $\theta$ is
the angle between the interatomic vector and the applied magnetic
field. In the actual experiments,\cite{mourachko,anderson} there
is a spatially varying magnetic field in the magneto-optical trap,
and the full analysis is much more complicated.

In the simplest case these waves are dispersionless and their
frequency, given in Eqs.~(\ref{Uk}) and (\ref{omegak}), depends on
the angle $\theta_{k}$ between the propagation vector and the
field's direction according to $1 - 3\cos^{2}\theta_{k} = -2
P_{2}\left(\cos\theta_{k}\right)$. These waves have not (to our
knowledge) been previously recognized, perhaps because their
occurrence is confined to systems that can be described by the
highly simplified Eq.~(1a), which additionally is treated in the
continuum approximation.

We remark that the continuum approximation becomes exact in the
long wavelength limit and is expected to have a wide range of
validity because the dipole-dipole interaction is long-ranged. The
polarization waves also exist on a regular lattice, with a
dispersion relation that can be calculated\@.\cite{cohen,wolynes}
In a frozen gas the excitation spectrum will spread out, but
localization effects are probably small, again because of the
long-range nature of the dipole-dipole interaction.

Although in our model system almost all atoms are in an excited
$s$ state, they could just as well be in the ground state. What is
important to the model is that this state couples effectively to a
single $p$ state. This is a good approximation for the Rydberg
atoms of the system we are modeling,\cite{mourachko,anderson} but
it is not a good approximation in general. Also, the size of the
dipole matrix element $\mu$ is particularly large for Rydberg
atoms.

Finally, we mention that the $1/r^{3}$ dipole-dipole interaction
does not include retardation effects. However, these are expected
to influence the results only by inducing some dispersion at very
long wavelengths.

In this paper we consider a particular mechanism of excitation of
the polarization waves, namely, we allow the resonant transfer
$ss^{\prime}\rightarrow pp^{\prime}$ to create the initial $p$
state that then interacts with the $s$ states\@.\cite{anderson}
However, other mechanisms of creation of the $p$ state would work
just as well as long as a single $p$ state is created. We have
also assumed that if only a few $p$ atoms are created, then each
can be treated independently in accordance with the sparse limit.

We outline (but do not carry out in detail in this paper) the
calculation of the transient behavior involving the excitation of
the newly-found polarization waves in a frozen gas. While this
allows us to establish contact with existing or currently planned
experiments, one should perhaps think instead of more direct ways
of observing the polarization waves.

\acknowledgments Acknowledgment is made to the Thomas F.\ and Kate
Miller Jeffress Memorial Trust for the support of this research.
We also wish to thank P. G. Wolynes for bringing to our attention
related work done by him and his collaborators, and T. Gallagher
for many useful discussions.

\begin{references}
\bibitem{footnote1} This result can also be obtained from
Eqs.~(\ref{Meqn}) and (\ref{Mexp}). Note that the limit
$U\rightarrow 0$, or $\mu\rightarrow 0$, corresponds to
$z\rightarrow\infty$.
\end{references}
\end{document}